\documentclass[prl,twocolumn,showpacs,superscriptaddress]{revtex4}

\usepackage{amsmath}
\usepackage{amsfonts}
\usepackage{graphicx}% Include figure files
\usepackage{dcolumn}% Align table columns on decimal point
\usepackage{bm}% bold math
\usepackage[usenames,dvipsnames]{color}
\usepackage{comment}
\usepackage{placeins}

\newcommand{\ket}[1]{\vert#1\rangle}
\newcommand{\bra}[1]{\langle#1\vert}
\def\proj#1#2{\ket{#1}\bra{#2}}
\def\opone{\leavevmode\hbox{\small1\kern-3.8pt\normalsize1}}

\begin{document}

\title{Spectral multiplexing for scalable quantum photonics using an atomic frequency comb quantum memory and feed-forward control} 

\author{Neil~Sinclair}
\affiliation{Institute for Quantum Science and Technology, and Department of Physics \& Astronomy, University of Calgary, 2500 University Drive NW, Calgary, Alberta T2N 1N4, Canada}
\author{Erhan~Saglamyurek}
\affiliation{Institute for Quantum Science and Technology, and Department of Physics \& Astronomy, University of Calgary, 2500 University Drive NW, Calgary, Alberta T2N 1N4, Canada}
\author{Hassan~Mallahzadeh}
\affiliation{Institute for Quantum Science and Technology, and Department of Physics \& Astronomy, University of Calgary, 2500 University Drive NW, Calgary, Alberta T2N 1N4, Canada}
\author{Joshua~A.~Slater}
\affiliation{Institute for Quantum Science and Technology, and Department of Physics \& Astronomy, University of Calgary, 2500 University Drive NW, Calgary, Alberta T2N 1N4, Canada}
\author{Mathew~George}
\altaffiliation{Present address: Department of Physics, CMS College, Kottayam 686 001 , India}
\affiliation{Department of Physics - Applied Physics, University of Paderborn, Warburger Stra{\ss}e 100, 33095 Paderborn, Germany}
\author{Raimund~Ricken}
\affiliation{Department of Physics - Applied Physics, University of Paderborn, Warburger Stra{\ss}e 100, 33095 Paderborn, Germany}
\author{Morgan~P.~Hedges}
\altaffiliation{Present address: Department of Physics, Princeton University, Jadwin Hall, Princeton, New Jersey 08544, USA}
\affiliation{Institute for Quantum Science and Technology, and Department of Physics \& Astronomy, University of Calgary, 2500 University Drive NW, Calgary, Alberta T2N 1N4, Canada}
\author{Daniel~Oblak}
\affiliation{Institute for Quantum Science and Technology, and Department of Physics \& Astronomy, University of Calgary, 2500 University Drive NW, Calgary, Alberta T2N 1N4, Canada}
\author{Christoph~Simon}
\affiliation{Institute for Quantum Science and Technology, and Department of Physics \& Astronomy, University of Calgary, 2500 University Drive NW, Calgary, Alberta T2N 1N4, Canada}
\author{Wolfgang~Sohler}
\affiliation{Department of Physics - Applied Physics, University of Paderborn, Warburger Stra{\ss}e 100, 33095 Paderborn, Germany}
\author{Wolfgang~Tittel}
\altaffiliation{Corresponding author. wtittel@ucalgary.ca}
\affiliation{Institute for Quantum Science and Technology, and Department of Physics \& Astronomy, University of Calgary, 2500 University Drive NW, Calgary, Alberta T2N 1N4, Canada}

\begin{abstract}
{
Future multi-photon applications of quantum optics and quantum information science require quantum memories that simultaneously store many photon states, each encoded into a different optical mode, and enable one to select the mapping between any input and a specific retrieved mode during storage. Here we show, with the example of a quantum repeater, how to employ spectrally-multiplexed states and memories with fixed storage times that allow such mapping between spectral modes. Furthermore, using a Ti:Tm:LiNbO$_3$ waveguide cooled to 3 Kelvin, a phase modulator, and a spectral filter, we demonstrate storage followed by the required feed-forward-controlled frequency manipulation with time-bin qubits encoded into up to 26 multiplexed spectral modes and 97\% fidelity.}
\end{abstract}


\maketitle

Further advances towards scalable quantum optics \cite{nphotfocus2009a,scully1997a} and quantum information processing \cite{kok2007a,sangouard2011a} rely on joint measurements of multiple photons that encode quantum states (e.g. qubits) \cite{kok2007a,sangouard2011a, nunn2013a}. However, as photons generally arrive in a probabilistic fashion, either due to a probabilistic creation process or due to loss during transmission, such measurements are inherently inefficient. For instance, this leads to exponential scaling of the time required to establish entanglement, the very resource of quantum information processing, as a function of distance in a quantum relay \cite{collins2005a}. This problem can be overcome by using quantum memories, which are generally realized through the reversible mapping of quantum states between light and matter \cite{lvovsky2009a, bussieres2013a}. For efficient operation, these memories must be able to simultaneously store many photon states, each encoded into a different optical mode, and subsequently (using feed-forward) allow selecting the mapping between input and retrieved modes (e.g., different spectral or temporal modes). This enables making several photons arriving at a measurement device indistinguishable, thereby rendering joint measurements deterministic. 
For instance, revisiting the example of entanglement distribution, a quantum relay supplemented with quantum memories changes it to a repeater and, in principle, the scaling from exponential to polynomial \cite{sangouard2011a, briegel1998a}.

Interestingly, for such multimode quantum memories to be useful, it is not necessary to map any input mode onto any retrieved (output) mode, but it often suffices if a single input mode, chosen once a photon is stored, can be mapped onto a specific output mode (e.g. characterized by the photon's spectrum and recall time) \cite{sangouard2011a,simon2007a}. This ensures that the photons partaking in a joint measurement, each recalled from a different quantum memory, are indistinguishable, as required, e.g., for a Bell-state measurement. We emphasize that it does not matter if the device used to store quantum states also allows the mode mapping, or if the mode mapping is performed after recall using appended devices -- we will refer to the system allowing storage and mode mapping as \emph{the memory}. 

To date, most research assumes photons arriving at different times at the memory (i.e. temporal multiplexing), and recall on demand in terms of variable storage time \cite{lvovsky2009a,bussieres2013a}. Here we show, with the example of a quantum repeater, that it is also possible to employ spectrally-multiplexed states and storage devices with fixed storage times, supplemented with frequency shifts based on feed-forward control. Furthermore, we report measurements using a highly broadband solid-state memory \cite{saglamyurek2011a} that demonstrate the required mapping between input and output modes with time-bin qubits encoded into up to 26 spectral modes and a fidelity of 0.97, thereby significantly violating the classical bound of 2/3 \cite{massar1995a}.

It is worth noting that for applications requiring short storage times, such as in linear optics quantum computing, a low-loss fibre could be sufficient to delay photons until a feed-forward signal arrives. However, for applications, such as a quantum repeater, in which storage times exceed around 10 $\mu$s, fibre transmission drops below 90\% and hence quantum state storage based on light-matter interaction will be necessary. Additionally, light-matter interaction affords more flexibility to perform processing tasks other than delaying \cite{saglamyurek2013a}.

Much theoretical and experimental work aiming at the development of quantum memory has been reported over the past decade \cite{lvovsky2009a,bussieres2013a,simon2010a}, and most criteria required for such a memory to be suitable for the aforementioned applications have been independently met. However, at most two (entangled) qubits have been stored simultaneously in a way that allowed selecting the mode mapping \cite{dai2012a}, and the scalability of the approach, which relied on encoding information into four spatial modes, to tens or hundreds of qubits and modes remains to be proven (we note related work by Lan et al. \cite{lan2009a} that, however, is not based on memories as defined above).

Rare-earth-ion doped crystals cooled to cryogenic temperatures have demonstrated to be promising storage materials, and many benchmark results have been reported \cite{saglamyurek2011a,clausen2011a,hedges2010a,usmani2010a,sabooni2013a,timoney2013a,jin2013a}. We emphasize that, when such crystals are used in conjunction with the atomic frequency comb (AFC) protocol, the independence of the multimode (i.e. multi-photon) capacity on optical depth constitutes an important advantage compared to other protocols \cite{afzelius2009a}. However, choosing the time of recall using control lasers to perform the mode mapping in the storage device is challenging with an AFC memory \cite{timoney2013a,afzelius2009a}. 

\begin{figure*}
\begin{center}
\includegraphics[width=2\columnwidth]{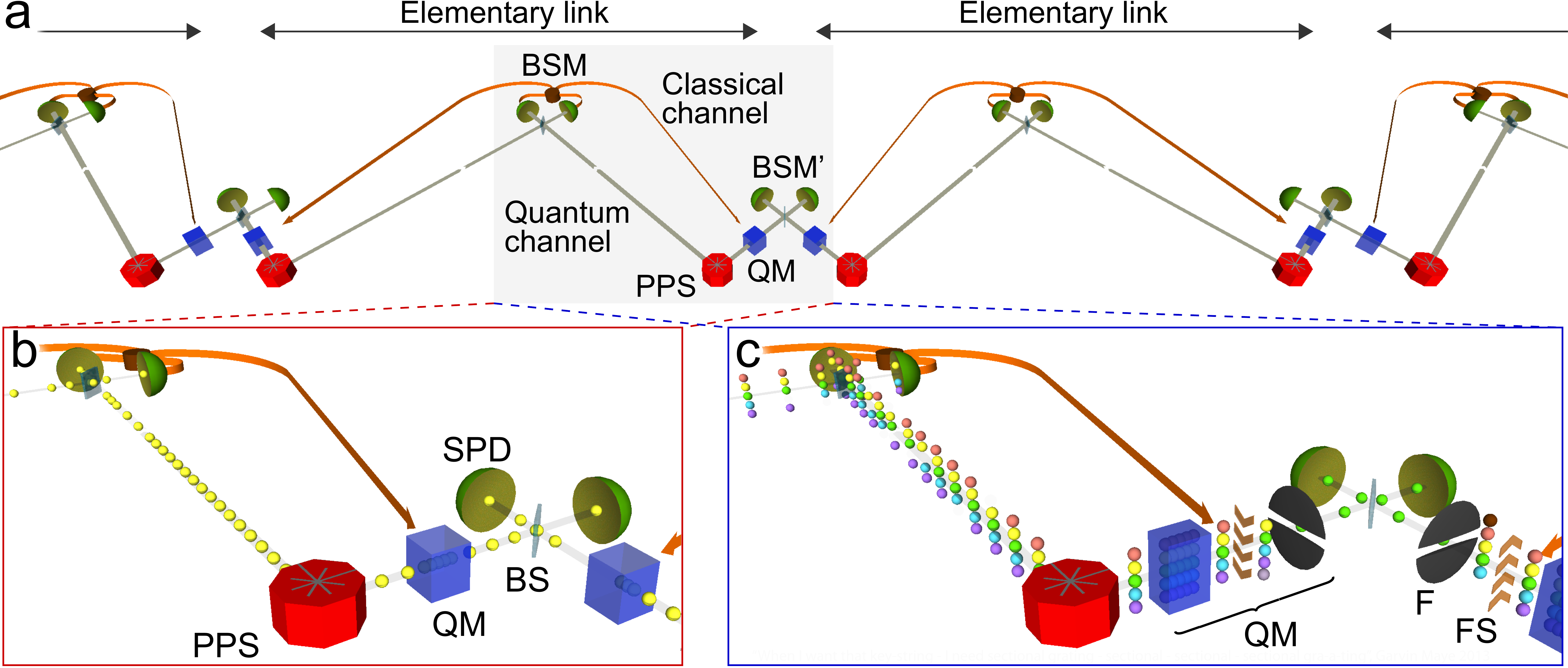}
\caption{(color online) Quantum repeater. 
\textbf{a}, Block diagram of a section of a quantum repeater that does not employ qubit multiplexing. A source generating entangled pairs of photons (PPS) is located at the end-point of each elementary link (i.e. node). One member per pair is stored in a quantum memory (QM), and the second member is sent over a 'quantum channel' to the centre of the link where it meets a member of an entangled pair generated at the other end of this link. The two photons' joint state undergoes a Bell-state measurement (BSM) -- comprised of a beam splitter (BS) and two single photon detectors (SPDs), and the result is communicated over a 'classical channel' back to the end-points to herald the establishment of entangled quantum memories by means of entanglement swapping \cite{sangouard2011a, briegel1998a,bennett1993a}. Entanglement is stored until the two memories that are part of an adjacent link, are also entangled. Then, photons are recalled from neighboring memories and subjected to BSM'. This results in the establishment of entanglement across the two links, and, by continuing this procedure with other links, entanglement is established between the end-points of the entire channel.
\textbf{b (c)}, Operation of a repeater node assuming temporal (spectral) multiplexing. Members of entangled photon pairs, each featuring the same spectrum (temporal profile and arrival time) but separated in time (frequency), are simultaneously stored in multimode quantum memories. A heralding (feed-forward) signal, derived from a successful BSM at the centre of each elementary link, indicates which of the stored photons is to be used for the remaining step of the protocol. The heralded photons are then recalled from adjacent memories such that they arrive indistinguishably at the BSM'. For temporal multiplexing, memories that allow adjusting the recall time as well as time-resolved detection are required, while for spectral multiplexing, the memories must incorporate adjustable frequency shifts (FS) and spectral filtering (F), and the BSM must distinguish different frequency bins.} 
 \label{fig:repeater}
\end{center}
\end{figure*}
\begin{figure}
\begin{center}
\includegraphics[width=\columnwidth]{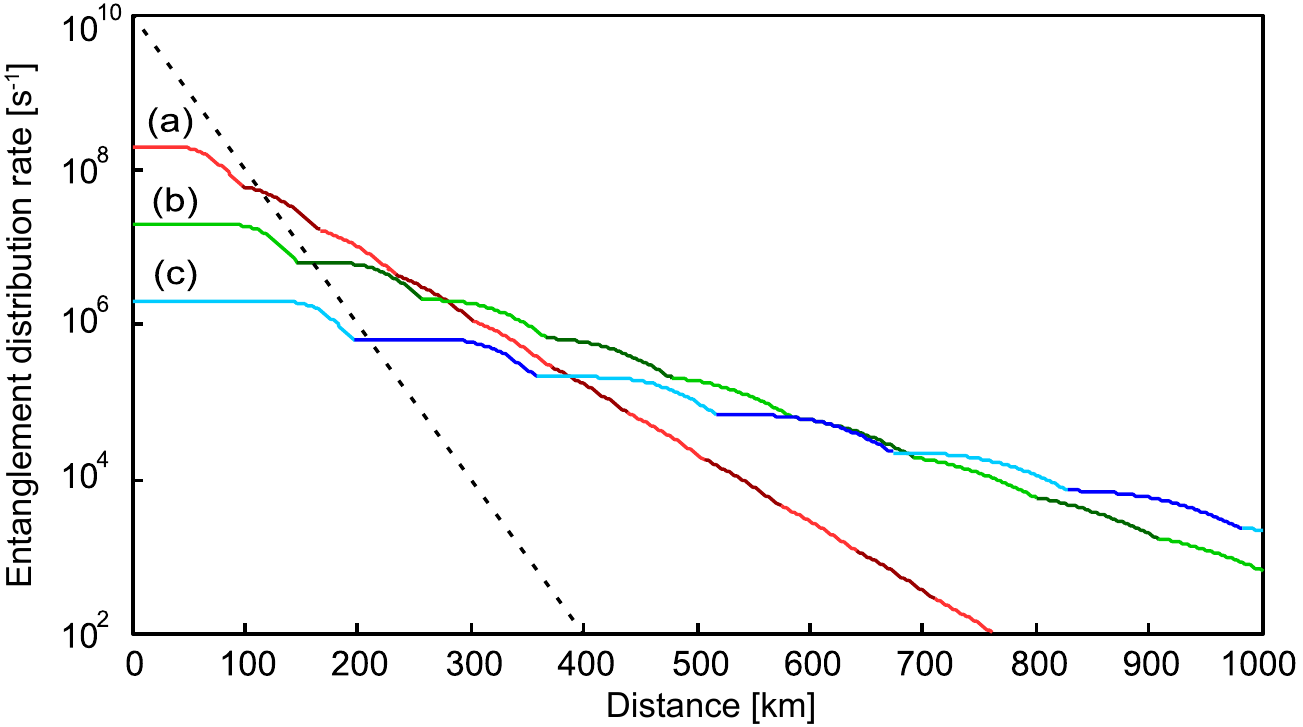}
\caption{(color online) Simulation of spectrally multiplexed quantum repeater performance. Optimal average entanglement distribution rate as a function of total distance. We assume loss of 0.2~dB/km, maximally entangled photon pairs emitted with $90\%$ probability per attempt, quantum memories with 90\% efficiency and total storage bandwidth of 300 GHz, and single-photon detectors with 90\% efficiency and negligible dark counts. Bi-coloured curves -- where a change in shading indicates the addition of an elementary link -- represent \textbf{(a)} $10^2$ (shown in red), \textbf{(b)} $10^3$ (shown in green), and \textbf{(c)} $10^4$ (shown in blue) spectral modes. The dotted line represents the direct transmission of members of entangled photon pairs produced at 10~GHz.}
 \label{fig:simulation}
\end{center}
\end{figure}

Drawing from the well-known temporal multiplexing approach, Fig.~\ref{fig:repeater} shows, with the example of a quantum repeater, how spectrally multimode quantum memories, including frequency shifters and filters, allow rendering photons indistinguishable without the need for a variable storage time. While a repeater that employs temporal multiplexing assumes all qubits to feature the same spectrum but to arrive at different times at the memory, our new approach assumes all qubits to arrive at the same time, but to feature distinct spectra (i.e. to be encoded into different frequency bins). The retrieval of a desired qubit at a given time and with a given spectrum can then be achieved by retrieving all qubits after the same storage time, selecting the shift of the spectra of all qubits such that the desired qubit occupies a previously agreed-upon frequency bin, and rejecting all other qubits using a filtering cavity. To quantify the performance of a quantum repeater based on spectral multiplexing, we calculate the average rate of successful distributions of entangled photon pairs over a lossy channel as a function of total distance. The results, shown in Fig.~\ref{fig:simulation}, show that useful performance can already be achieved with 100 spectral modes, which is clearly feasible in near future. Further information regarding the derivation of these results, and comparison with the temporal multiplexing scheme are contained in the Supplemental Material.

\begin{figure}
\begin{center}
\includegraphics[width=\columnwidth]{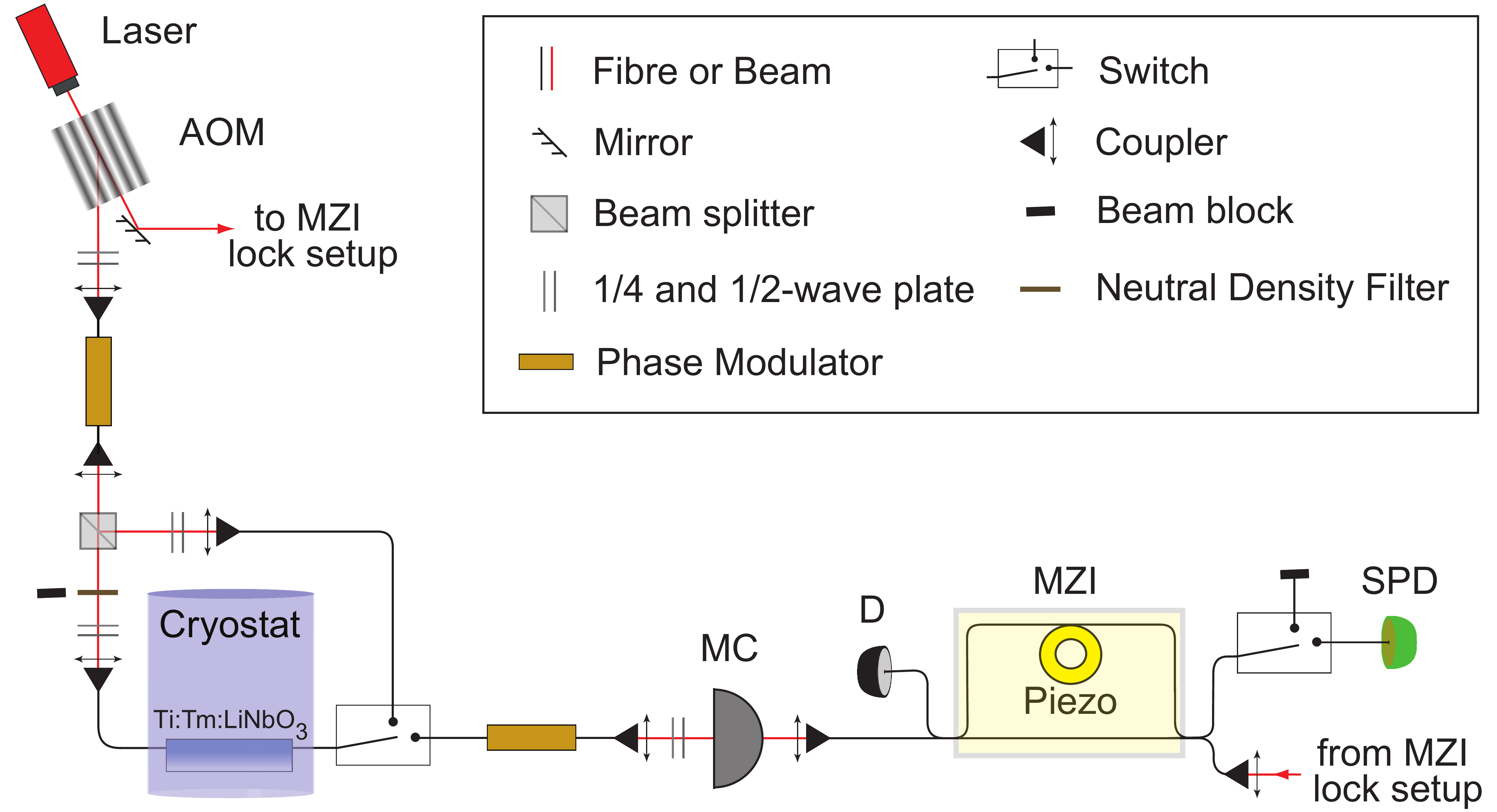}
\caption{(color online) Schematics of the experimental set-up. The output of a frequency-stabilized continuous-wave laser at 795.4 nm wavelength is amplitude modulated with an AOM and serrodyne chirped \cite{johnson1988a} over disjoint frequency intervals using a phase modulator (PM). During 5~ms the laser light creates a broadband multimode AFC (see Fig.~\ref{fig:26modeprobes}) in a Tm:Ti:LiNbO$_3$ waveguide located inside a 3~Kelvin cryostat and exposed to a magnetic field of 88 Gauss \cite{saglamyurek2011a, sinclair2010a}. After a 2~ms wait time, during the next 5 ms, the AOM generates, with 4~MHz repetition rate, up to 26 spectrally multiplexed pairs of 15~ns-long Gaussian-shaped pulses (pulses in the pairs are separated by 20~ns), whose relative phases and central frequencies are set using the PM. The subsequent attenuator, or beam block, reduces the mean number of photons per pulse pair to 0.5, 0.1, or zero, respectively. The resulting time-bin qubits are then sent into the waveguide, and stored for 60~ns. Frequency-selective recall is achieved by means of a second PM, combined with a monolithic cavity (MC) having 70~MHz line-width \cite{palittapongarnpim2012a}. Finally, 
the recalled photons are detected using a Si-avalanche photodiode-based single photon detector (SPD) (allowing projections onto $|e\rangle$ and $|l\rangle$), or a phase-stabilized Mach-Zehnder interferometer (MZI) followed by a SPD (allowing projections onto $\tfrac{1}{\sqrt{2}}\big(|e\rangle \pm |l\rangle\big)$).}
\label{fig:setup}
\end{center}
\end{figure}

\begin{figure}
\begin{center}
\includegraphics[width=\columnwidth]{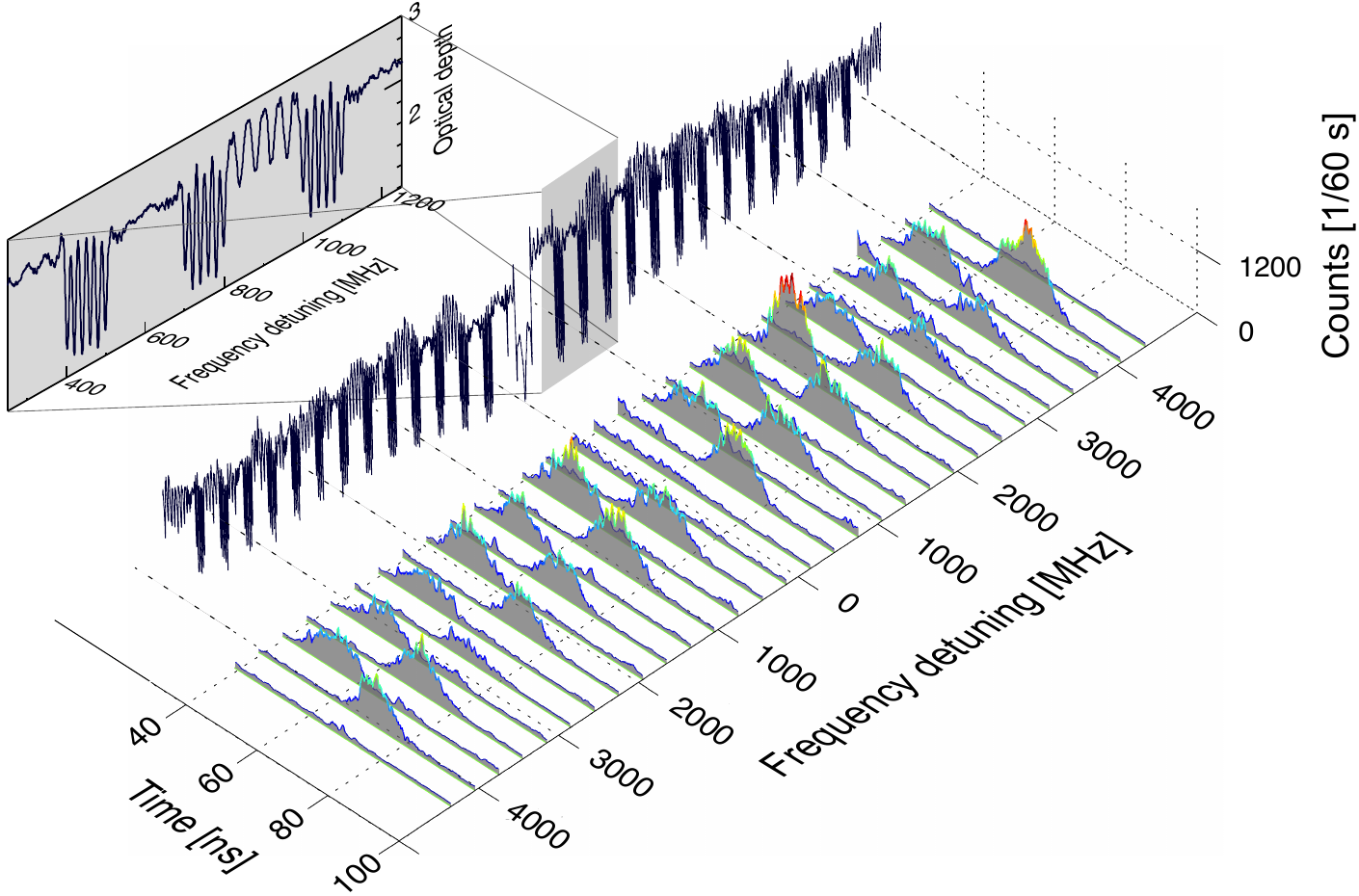}
\caption{(color online) Multimode storage and frequency selective recall. Histogram of arrival times of 26 simultaneously stored qubits, each containing 0.5 photons on average. Qubits are prepared in separate spectral modes and alternating temporal modes (i.e. $\ket{e}$ and $\ket{l}$), and are each recalled individually. The cavity resonance was set to 200~MHz detuning. No recall of qubits is observed in spectral modes at $\pm 150$ and $\pm 4350$ MHz detunings where no AFCs were prepared. The back panel and inset show the multi-binned AFC absorption profile utilized. Modulation outside of the individual combs is due to higher order effects from the phase modulation.}
\label{fig:26modeprobes}
\end{center}
\end{figure}

Conjecturing similarly promising results for other multi-photon applications, we now experimentally characterize the feasibility of multimode storage and feed-forward-controlled read-out in the frequency domain. A schematic of our setup is depicted in Fig.~\ref{fig:setup}. It performs four tasks: First, to prepare the memory, laser light is temporally and spectrally modulated, and then sent into a Ti:Tm:LiNbO$_3$ waveguide \cite{saglamyurek2011a, sinclair2010a}, thus spectrally tailoring the inhomogeneous absorption line of thulium into a series of equally-spaced absorption peaks -- an AFC. For multimode storage, the preparation procedure is repeated at different detunings with respect to the original laser frequency, resulting in twenty-six, 100 MHz-wide AFC's that are, with the exception of the region around zero detuning, spectrally separated by 200 MHz gaps (see Fig.~\ref{fig:26modeprobes}). 
Second, our setup simultaneously generates many time-bin qubits of the form $\ket{\psi}=\alpha\ket{e}+\beta\ket{l}$, encoded into single-photon-level, phase randomized laser pulses of different intensities, in up to 26 frequency bins. Here, $|\alpha |^2+|\beta |^2=1$, and $\ket{e}$ and $\ket{l}$ describe early or late emitted laser pulses, respectively. Third, the qubits are sent into the waveguide memory, where the absorption of each photon occupying a specific frequency bin leads to a collective excitation shared by the atoms forming the corresponding AFC. After a preset storage time $T_{s}=1/\Delta$ (where $\Delta$ is the AFC peak spacing), the photons are emitted in their original state and spectral mode \cite{afzelius2009a}. For selecting the recalled mode, the spectra of all simultaneously recalled photons are frequency shifted using another phase modulator \cite{johnson1988a}, and all but the desired photons are rejected using a filter cavity with fixed resonance frequency \cite{palittapongarnpim2012a}. 
Finally, projection measurements onto time-bin qubit states $\ket{e}$ or $\ket{l}$, or $(\ket{e}\pm\ket{l})/\sqrt{2}$ are performed. As we describe in detail in the Supplemental Material, we post-process the measured data to assess a key figure of merit -- the lower bound on the storage fidelity $\mathcal{F}_L^{(1)}$ --  only from laser pulses containing exactly one photon. This procedure justifies the use of attenuated laser pulses instead of single photons to encode qubits for the purpose of our investigation. Further details about the AFC preparation, qubit generation, measurements and fidelity calculations, as well as current limitations resulting in a $1.5\times10^{-4}$ overall memory efficiency are contained in the caption of Fig.~\ref{fig:setup} and the Supplemental Material.

In the first experiment we simultaneously store 26 qubits, alternating between $\ket{e}$ and $\ket{l}$, each prepared in one of the 26 spectral bins containing AFCs. All qubits are recalled after 60~ns, and subsequently frequency shifted and spectrally filtered. Fig.~\ref{fig:26modeprobes} shows histograms of detections as a function of time for 30 different frequency shifts, for which the cavity filtering is expected to select the recall of at most one qubit. 
Our results indicate that, with little cross-talk from the directly neighboring frequency bins, we can simultaneously store many qubits featuring disjoint spectra, and recall each qubit individually. We note that the total storage bandwidth of Tm:LiNbO$_3$ exceeds 300~GHz \cite{sun2012a}, which, in principle, allows expanding the current AFC to comprise more than 1000 spectral bins.

Next, to further examine the effect of cross-talk between spectral modes, we first store and retrieve a 'test' qubit prepared in $\ket{l}$ in the spectral bin having 1350 MHz detuning (with vacuum in all other spectral bins). We shift the test qubit into cavity resonance, and measure the probability to detect it in an early or a late temporal mode, which allows calculating the fidelity $\mathcal{F}_l$  of the recalled state with the input state (here and henceforth, the subscript index indicates the qubit's originally prepared state). We then increase the number of simultaneously stored qubits by creating them in neighboring spectral bins, and repeat the fidelity measurement with the test qubit. Note that all additional qubits are prepared in the orthogonal $\ket{e}$ state, such that the reduction of the fidelity of the test qubit due to cross-talk is maximized. The result, further described in the Supplemental Material, shows that cross-talk (due to the Lorentzian-shaped cavity resonance line) is restricted to qubits separated by at most two frequency bins.

\begin{table}
\centering
\renewcommand{\arraystretch}{1.5}
\begin{tabular}{ c  c  c }
\hline
\hspace{1mm} Photon input \hspace{1mm} & $\mathcal{F}_{e/l}$ & $\mathcal{F}_{+/-}$  \\
\hline\hline
$\mu=0.5$ & \hspace{1mm} $(94.67 \pm 0.43)\%$ \hspace{1mm} & \hspace{1mm} $(94.52 \pm 0.67)\%$ \hspace{1mm} \\  
$\mu=0.1$ & $(91.56 \pm 1.35)\%$ & $(85.14 \pm 2.73)\%$ \\ 
$n=1$ & $(94.03 \pm 1.87)\%$ & $(97.76 \pm 5.54)\%$ \\ 
\hline
\end{tabular}
\caption{Storage and recall fidelities, $\mathcal{F}_{e/l}$ and $\mathcal{F}_{+/-}$, for test qubits encoded into attenuated pulses of mean photon number $\mu$, and lower bounds $\mathcal{F}^{(1)}_{L,e/l}$ and $\mathcal{F}^{(1)}_{L,+/-}$ on storage and recall fidelities for qubits encoded into single-photon states ($n=1$) derived using decoy state analysis \cite{ma2005a}. One-standard-deviation uncertainties are calculated from statistical uncertainties of photon counts.}
\label{table:5modeZX}
\end{table}

Finally, we quantify the storage and recall fidelity for arbitrary qubit states stored in the AFC with multiple spectral bins shown in Fig.~\ref{fig:26modeprobes}. Supported by the previous result, we create and simultaneously store time-bin qubits prepared in five spectral bins located between 750 and 1950~MHz detuning. A test qubit in state $\ket{\psi}\in [ \ket{e}, \ket{l}, \frac{1}{\sqrt{2}}(\ket{e}+\ket{l}), \frac{1}{\sqrt{2}}(\ket{e}-\ket{l}) ]$ is prepared in the central bin (at 1350~MHz detuning), and, for the reason already described above, the qubits in the four neighboring bins are prepared in the orthogonal state. We set the frequency shift to recall only the test qubit, and calculate the fidelity with its original state. This measurement is performed with mean photon numbers per qubit of 0.5, 0.1 and zero. Each measurement is taken over 60 s and the cavity resonance was set to a detuning of 3~GHz. 

The resulting fidelities $\mathcal{F}_{e/l}$ and $\mathcal{F}_{+/-}$, averaged over each set of basis vectors (e.g. $\mathcal{F}_{e/l}=\frac{1}{2}(\mathcal{F}_e+\mathcal{F}_l)$), for mean photon numbers of 0.5 and 0.1 are displayed in Table ~\ref{table:5modeZX}. In addition, the table shows the lower bounds on the fidelities that we would have obtained if, with no other things changed, we had performed our experiments with qubits encoded into individual photons. These bounds, denoted by $\mathcal{F}^{(1)}_{L,e/l}$ and $\mathcal{F}^{(1)}_{L,+/-}$, are derived using a decoy state method that underpins the security of quantum key distribution based on attenuated laser pulses (further explanation of this method is found in the Supplemental Material and \cite{ma2005a}). We find that all fidelities exceed the maximum value of 2/3 achievable using a classical memory \cite{massar1995a}. Deviations from unity fidelity are due to the limited frequency shift efficiency of the phase modulator, limited suppression of the cavity, limited visibility and stability of the Mach-Zehnder interferometer used for certain projection measurements, and remaining laser frequency and power fluctuations. Furthermore, the measurements with mean photon number of 0.1 are impacted by system loss and detector dark counts. Finally, by averaging the single-photon fidelities over all (properly weighted) input states, we derive our key figure of merit -- the lower bound on the single-photon fidelity $\mathcal{F}^{(1)}_{L}= \frac{1}{3}\mathcal{F}^{(1)}_{L,e/l}+\frac{2}{3}\mathcal{F}^{(1)}_{L,+/-} = 0.97 \pm 0.04$. It exceeds the classical bound by 7.5 standard deviations, proving our memory to be suitable for applications of quantum optics and quantum information science.

In conclusion, we have shown for the first time that it is possible to combine the simultaneous storage of multiple qubits with feed-forward controlled mapping between input and output modes using a protocol that allows scaling the number of qubits to many hundreds. This is likely to accelerate the development of quantum repeaters, linear optics quantum computing, and advanced quantum optics experiments, in particular if our frequency-based approach is combined with multiplexing using other degrees of freedom. For instance, considering as few as 10 frequency, 10 temporal and 10 spatial modes, photons in 1000 different modes can be multiplexed, which already suffices for a quantum repeater. Or, considering 500 frequency, 10 spatial \cite{lan2009a} and 400 temporal modes \cite{bonarota2011a}, one could simultaneously store 10$^6$ qubits. Note that any multiplexed degree of freedom can be manipulated to render photons indistinguishable -- in our demonstration we used frequency.

\acknowledgments
This work is supported by Alberta Innovates Technology Futures (AITF), the National Sciences and Engineering Research Council of Canada (NSERC), the US Defense Advanced Research Projects Agency (DARPA) Quiness Program under Grant No. W31P4Q-13-1-0004, the Killam Trusts, and the Carlsberg Foundation. The authors thank Thierry Chaneli\`ere for help with the laser locking system, Vladimir Kiselyov for technical support during various stages of the experiment, and Alex Lvovsky for lending us the filtering cavity.\\

\newpage

\begin{center}
	\textbf{SUPLEMENTAL MATERIAL}
\end{center}

\section{A: Performance of spectrally multiplexed quantum repeater}

In the main text we briefly described an approach to quantum repeaters based on multiplexing in the spectral domain. We consider quantum memories with pre-set storage times, as discussed in detail in the main text. In this section we will first describe our architecture in more detail and then describe some of the calculations used to generate Fig. 2 in the main text: the average rate of successful entanglement creation across a lossy quantum channel.

In our architecture there exists a lossy quantum channel with length $L$ and loss coefficient of $\alpha$. We split this channel into $n$ \emph{elementary links}, each of length $L/n$. The goal of our architecture is, first, to nearly d{eterministically create entanglement between the ends of each elementary link, which we will refer to as \emph{elementary-level entanglement}, and, second, create entanglement between the ends of the entire quantum channel via entanglement swapping between the elementary links. The number of elementary links $n$ is then optimized based on the length and loss of the channel as well as the performance of other elements used in the architecture, such as detectors and quantum memories, to maximize the rate of entanglement creation.  To deterministically create elementary entanglement we place a spectrally-multimode quantum memory at each end of the elementary link, as seen in Fig. 1 of the main text. Immediately beside each quantum memory we place a source that generates two-photon entanglement in many spectral modes simultaneously (i.e. a spectrally-multimode two-photon entanglement source). Note that the two-photon entangled state in each mode is independent of all other spectral modes (e.g. the source may generate the maximally-entangled $\ket{\psi-}$ Bell state in each spectral mode: $\ket{\Psi-} = \ket{\psi-}_{\omega_1}\ket{\psi-}_{\omega_2}...\ket{\psi-}_{\omega_m}$) and that each source must use the same, pre-agreed upon, set of spectral modes. One photon from each mode on each side of the elementary link is immediately stored in the quantum memory while the second photon of each pair travels to the centre of the elementary link.

At the centre of the elementary link is a \emph{centre station} that attempts a Bell-state measurement (BSM) on each spectral mode, using one photon from each source. In general the BSM is probabilistic due to channel loss, limited detector efficiencies, the use of linear optics for the BSM~\cite{lutkenhaus1999a} and loss in other optical elements. However, as photons were emitted in many spectral modes simultaneously, the probability that at least one mode yields a successful BSM can be made arbitrarily close to $1$ given a sufficiently large number of spectral modes. In essence, a successful BSM in spectral mode $j$ heralds elementary entanglement in mode $j$ between the quantum memories (multiple simultaneous BSM successes are ignored for simplicity). The centre station sends a classical feed-forward signal informing the quantum memories of the spectral mode in which they now share entanglement. 

Finally, to establish entanglement between the end points of the entire quantum channel, neighbouring elementary links must perform entanglement swapping, involving a BSM' at the intersection of neighbouring elementary links -- where the prime serves to distinguish this BSM from the one performed at the centre station.  As neighbouring links likely have elementary entanglement in different spectral modes, a frequency conversion is necessary before the entanglement swapping BSM'.  Then, if the intersection of all elementary links successfully performs a BSM' (again probabilistic), entanglement is created between the end points of the quantum channel.

The calculation of the average rate of entanglement creation across the quantum channel can be broken down into three parts: The first part is to calculate the probability to create elementary entanglement across all $n$ single elementary link, denoted P(elementary). The second part is to calculate the probability to successfully create entanglement across the entire quantum channel, denoted P(success), and the third part is to calculate the time between successive attempts, whose inverse gives the rate of entanglement creation, denoted R(success). To begin with P(elementary), we discuss the case where all sources are probabilistic single-pair entanglement sources (e.g. a source that with probability $\rho$ emits $1$ pair of maximally entangled photons and with probability $(1-\rho)$ emits vacuum. Note that if one includes multi-pair emissions typical in entanglement sources based on spontaneous parametric downconversion, four-wave mixing, etc. then one will never create maximally entangled states across the channel. In which case, one must set some kind of tolerance on how much deviation from maximal entanglement is acceptable, such as requiring a fidelity over $90$\%. The operating conditions of the repeater architecture could then be optimized to meet this requirement, as done in~\cite{sangouard2011a} or in~\cite{khalique2013a} for the closely related quantum relay.

Considering a single elementary link, using only one spectral mode, the probability of a successful BSM at the centre station is given by
\begin{equation}
	\text{P({\small 1 mode})} = \dfrac{1}{2}\times  \left( \eta_{d1} \times \rho \times 10^{-\alpha L/2n } \right) ^2 \quad ,
\end{equation}
where $\eta_{d1}$ is the efficiency of the centre station's detectors, $\alpha$ is the loss coefficient of the channel and $L/2n$ is the length that each photon must travel. The square arises as two photons must travel $L/2n$ distance each and both must be detected. Furthermore, we have assumed that the detectors are noiseless, that the BSM is carried out with linear optics (i.e. is at most $50$\% successful and hence the pre-factor of $1/2$, although in principle this can be made $100$\% with non-linear optics), and that the visibility of the two-photon measurement is perfect.

Thus, if each source instead emits $m$ spectral modes simultaneously, the probability that all modes result in unsuccessful BSMs is given by
\begin{equation}
	\text{P({\small m modes, all fail})} = \left( 1 - \dfrac{1}{2}\left( \eta_{d1}\rho10^{-\alpha L/2n } \right) ^2 \right) ^m .
\end{equation}

In general, an extra device to distinguish the different spectral modes (e.g. wavelength-divsion-multiplexers, highly dispersive media, etc.) has to be used and will introduce extra loss. This could be included by appropriately lowering the detection efficiencies or by including another multiplicative term (within the innermost parentheses). Then, the probability that at least one spectral mode results in a successful BSM, which creates the elementary entanglement, is
\begin{multline}
	\text{P({\small m modes, not all modes fail})} = \\ 1-\left( 1 - \dfrac{1}{2}\left( \eta_{d1}\rho10^{-\alpha L/2n } \right) ^2 \right) ^m \ .
	\label{eq:mnofail}
\end{multline}
It is clear from Eq.~\ref{eq:mnofail} that by choosing the number of spectral modes $m$ sufficiently high, one can guarantee elementary entanglement with probability approaching one on every attempt.  Then, the probability that all $n$ elementary links successfully create entanglement is
\begin{equation}
	\text{P({\small elementary})} = \left( 1-\left( 1 - \dfrac{1}{2}\left( \eta_{d1}\rho10^{-\alpha L/2n } \right) ^2 \right) ^m \right) ^n.
\end{equation}

Secondly, once all elementary links have established elementary entanglement, entanglement swapping between the links is attempted. This involves recalling photons from the memories including the frequency shifting and filtering of all but one mode with total efficiency $\eta_{mem}$, and a BSM' ($50$\% efficiency) with inefficient detectors (with detection efficiency $\eta_{d2}$). Note that $\eta_{mem}$ is defined here as the probability that a photon is stored, is retrieved at the later time and is not lost during frequency shifting and filtering. If there are $n$ elementary links there are $(n-1)$ such entanglement swappings to perform, giving
\begin{multline}
	\text{P({\small m modes, n links, all links successful \& swapped})} = \\ \left(\dfrac{1}{2}\eta_{mem}^2\eta_{d2}^2\right) ^{(n-1)} \times \text{P(elementary)}.
	\label{eq:Pelementary}
\end{multline}

\begin{figure}
		\includegraphics[width=\columnwidth]{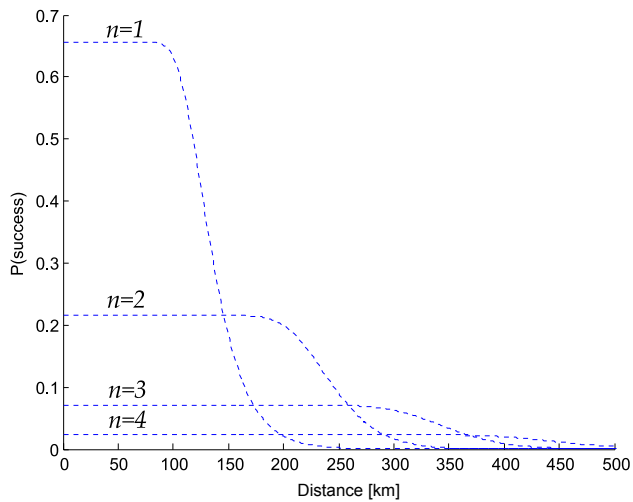}
		\caption{The probability to create entanglement across the entire channel per attempt, P(success), as a function of total distance, for reasonable values ($m=1000$~modes, $\eta_{mem} = \eta_{d1} = \eta_{d2} = \rho = 90$\% and $\alpha = 0.2$ dB/km). Each curve is for a different number of elementary links $n = 1 ... 4$.}\label{SIfig:Psuccess}\vspace{72pt}
\end{figure}
The entanglement is only useful if it can be recalled from the memories at the ends of the quantum channel and detected, and thus the probability, P(success), of generating entanglement across the entire quantum channel is given by Eq.~\ref{eq:Pelementary} $\times (\eta_{mem}^2\eta_{d2}^2)$ or (with minor simplification):
\begin{multline}
	\text{P({\small success})} = \\ \dfrac{ \left( \eta_{mem}\eta_{d2}\right) ^{2n}}{2 ^{n-1}} \ \times \ \left( 1-\left( 1 - \dfrac{1}{2}\eta_{d1}^2\rho^210^{-\alpha L/n } \right)^{\! m} \right)^{\! n}.
\end{multline}
Here, P(success) has been grouped into two terms corresponding to, first, the probability to successfully swap entanglement to connect elementary links and, second, P(elementary) -- the probability to create entanglement across all elementary links.  It is clear that all the ``loss" -- or deviation from deterministic success -- is contained within the the first term (connecting elementary links), while the second term is almost independent of loss, provided, again, that $m$ is sufficiently high. This is depicted in Fig.~\ref{SIfig:Psuccess} where we plot P(success) as a function of distance for $m = 1000$ and various $n$.  At low distance, for all $n$, P(success) is constant as $m$ is high enough to keep the second term close to $1$. Notice that in this flat region, for higher $n$, P(success) decreases. This is because using more elementary links requires more memories, detectors and BSMs, all with limited efficiency, which decreases the first term of P(success). For each $n$ there is a distance where P(success) drops sharply, which is where the second term begins to significantly drop below $1$ (i.e. $m$ is not sufficiently large to ensure elementary entanglement). However, for larger $n$, this sharp decrease occurs at a larger distance. This is because for larger $n$ (and a given total channel length $L$) the length of each elementary link is smaller and then P(elementary) $\approx 1$.
Thus, at larger distances it becomes advantageous to use more elementary links.
\begin{figure}
		\includegraphics[width=\columnwidth]{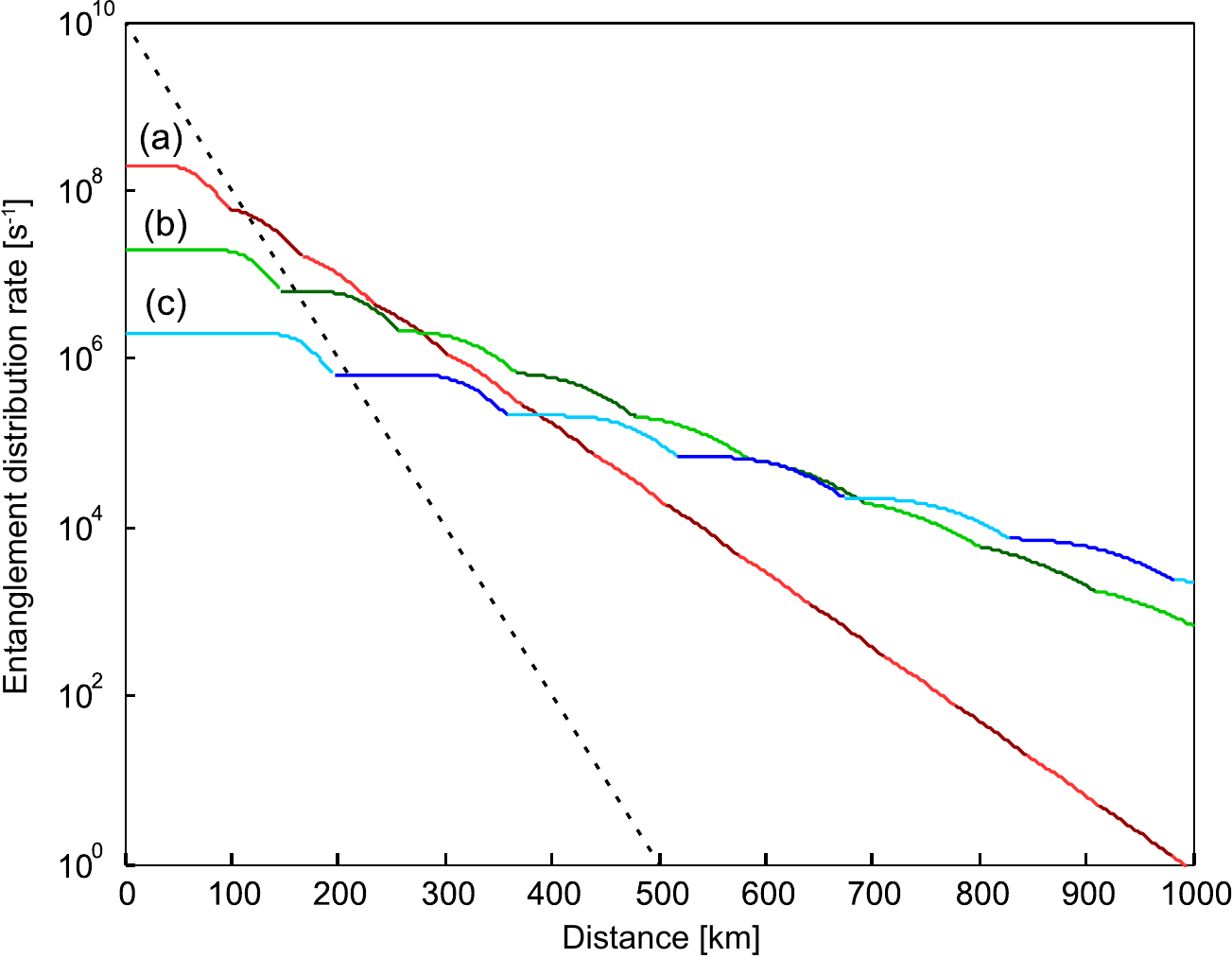}
		\caption{R(success) as a function of total distance, for reasonable values discussed in the text. Bi-coloured curves, where a change in colour indicates that the optimal value of $n$ has increased by one, represent (a) $10^2$ (shown in red), (b) $10^3$ (shown in green) and (c) $10^4$ (shown in blue) spectral modes. The dotted line represents direct transmission of members of entangled photon pairs produced at $10$ GHz.}\label{SIfig:Rsuccess}
\end{figure}

Lastly, we calculate the actual rate of entanglement creation R(success), which is equal to P(success) divided by the time between successive attempts. If the quantum memories are limited to storing only one qubit per spectral mode at a time, then the attempt rate would be limited by the communication time across the elementary link ($L/(nc)$, where $c$ is the speed of light), as quantum memories cannot use their elementary entanglement until they learn the result of the BSM at the centre station. However, if the quantum memories can store many qubits in each spectral mode simultaneously (along with $m$ spectral modes) then the attempt rate is limited by the device with the smallest operating bandwidth, assuming that all photons are Fourier-transform-limited (the spectral bandwidth of a device determines the shortest pulse duration the device can handle without loss, which determines the time between successive pulses). As optical fibre and other standard optical components already operate over many nanometers, and single-photon detectors with $10$~ns recovery time have been built~\cite{walenta2012a}, it is reasonable to assume that quantum memories will be the bandwidth-limiting device.  Assuming that quantum memories have a total bandwidth of $B$ and, again, that $m$ spectral modes will be used (which limits the bandwidth per mode) and allowing for a bandwidth inefficiency $w$ as many implementations will not be able to use the entire memory bandwidth, the time between successive attempts is given by $(w\times m)/B$. Therefore,
\begin{multline}
	\text{R({\small success})} = \\ \dfrac{B \ \times \ \left( \eta_{mem}\eta_{d2}\right) ^{2n} \ \times \ \left( 1-\left( 1 - \dfrac{1}{2}\eta_{dc}^2\rho^210^{-\alpha L/n } \right)^{\!\! m} \right)^{\!\! n}}{wm \times 2^{n-1}}.
\end{multline}

Knowing values for all parameters we can calculate the rate of entanglement generation for any distance and any number of elementary links $n$. We choose the $n$ that gives the highest rate and plot the entanglement generation rates in Fig.~\ref{SIfig:Rsuccess} (identical to Fig. 2 in the main text). We have performed this optimization for $m = 100$, $1000$ or $10000$ spectral modes using $B = 300$ GHz (according to the zero-phonon line of a Tm:LiNbO$_3$ crystal \cite{sun2012a}), $w = 10$, $\eta_{mem} = \eta_{d1} = \eta_{d2} = \rho = 90$\% and $\alpha = 0.2$ dB/km. In Fig.~\ref{SIfig:Rsuccess} we have also included the rate of success for direct transmission assuming a $10$ GHz pulsed entanglement source. By direct transmission we specifically envisage placing a source of two-photon entanglement at one end of the quantum channel and then sending one photon from the entangled pair across the quantum channel. If the travelling photon arrives at the other end then entanglement creation has been successful. This probability is simply R(success) = R(attempt)$\times10^{-\alpha L}$. Note that $w=10$ is reasonable for our implementation, given that AFC with rare-earth-ion doped crystals require shelving space (in the spectral domain) for ions (roughly a factor of $2$) and that time-bin qubits are composed of two temporally-separated pulses (roughly a factor of $4$). 

One can see that our architecture scales considerably better than direct transmission even for as little as $m=100$ spectral modes. At shorter distances, when channel loss is minimal, one sees that the rate of success is optimized with fewer spectral modes, as this allows for a higher attempt rate (more bandwidth per spectral mode allows for shorter pulses). However, at longer distances, more spectral modes become advantageous. This is because more spectral modes allows for longer elementary links, which decreases the main source of loss -- inefficient entanglement swapping between elementary links. In fact, the limited success probability for entanglement swapping between elementary links creates exponential scaling in distance for our architecture (albeit considerably better scaling than direct transmission), but this could be avoided with further advances. For instance, we assumed earlier that if elementary entanglement was established in multiple spectral modes on a single elementary link during an attempt, then only one pair was used for swapping. If every pair could be used for swapping \cite{collins2007a} then exponential scaling could be avoided if every attempt generated many pairs of elementary entanglement per attempt, which could be achievable by increasing $m$.

For comparison with temporal multiplexing architecture \cite{sangouard2011a}, it is evident that, with ideal resources, temporal multiplexing would perform similar to spectral multiplexing. 
This is because given a total available bandwidth $B$ one can subdivide it into $m$ spectral bins and store $m$ spectral modes of duration $t_m= m/B$ for spectral multiplexing. Conversely, one may utilize the entire bandwidth to store $m$ pulses (temporal modes) of duration $t_1=1/B$ during the same time $t_m$ for temporal multiplexing (for simplicity we have ignored the factor $w$, which impacts both protocols similarly). Hence, during a set time the two approaches afford the same number of attempts to succeed at entanglement swapping at the elementary link level, and therefore feature comparable performance. Yet, this is not true any more if we take material limitations into account.

For a more realistic comparison, let us take into account two key material properties of rare-earth ion doped materials. The first is the inhomogeneous broadening of the zero-phonon line, which determines the total available bandwidth. The second is the ground level splitting, which limits the bandwidth of the individual spectral bins. Tm:LiNbO$_3$ has a particularly large inhomogeneous broadening of 300 GHz and, as shown in Fig.~\ref{fig:highfin} in the Supplemental Material, we can achieve a splitting of the Zeeman ground levels of at least 500 MHz by application of a sufficiently large magnetic field. This is more than the largest bin width of 300 MHz used in our simulations (for $m=100$). Hence, the predicted repeater performance in the case of spectral multiplexing is not impeded by the properties of our storage material.

Realizing temporal multiplexing with recall on demand in AFC memories requires mapping of optically excited coherence onto long lived ground state coherence (often called spin-wave mapping – see  \cite{afzelius2010a,timoney2013a,gundogan2013a,jobez2014a} for recent progress). Please recall that the maximal bandwidth of a high-efficiency AFC is determined by the ground-level splitting, which is around 10~MHz in the materials used to date to implement the protocol (much less than 300~GHz, as needed for repeater performance comparable with that derived above for a spectral-multiplexing-based architecture). We point out that the value of 10~MHz is also utilized in \cite{sangouard2011a} to calculate the average entanglement distribution time for a temporal multiplexing protocol with $m=100$ modes and all other parameters as in our simulations. The simulations in \cite{sangouard2011a} show that the temporally multiplexed repeater out-performs a 10~GHz direct transmission protocol at around 510~km fibre length. From Fig.~\ref{SIfig:Rsuccess} in our Supplemental Material (also Fig. 2 in our main text) it is clear that our proposed architecture outperforms the temporal multiplexing scheme quite convincingly.
\\

\section{B: Preparation of the multimode atomic frequency comb (AFC)}

Each AFC, present over the bandwidth of each spectral bin, is comprised of a periodic modulation of the optical depth composing the 795 nm, inhomogeneously broadened $^3H_6\leftrightarrow {}^3H_4$ absorption line in Tm. It is generated by optically pumping ions to off-resonant nuclear Zeeman levels \cite{saglamyurek2011a,sinclair2010a}, and from troughs in the AFC to neighbouring peaks (as we will describe below, this approach limits the efficiency of the AFC memory to approximately $17$\%). Optical pumping is achieved by intensity and frequency modulating laser light via an AOM and phase modulator, respectively (refer to Fig. 3 in main text). We implement frequency sweeps by driving our phase modulator with sawtooth-shaped (i.e. serrodyne) waveforms originating from a 20 GS/s arbitrary waveform generator, and simultaneously drive the AOM with a 350 MHz signal when optical pumping is desired. The memory storage time $T_{s}$ is set by the AFC peak spacing $\Delta = 1/T_{s}$ between neighbouring teeth composing the AFC. In most experiments, the parameter $\Delta$ is set to 17 MHz, yielding 60 ns storage time. The only exception are the measurements leading to the results shown in Fig.~\ref{fig:highfin} where $\Delta=100$ MHz, yielding $T_{s} = 10$ ns.

To create a multi-spectral-binned (i.e. multimode) AFC, we program the phase modulator to quickly shift the laser frequency to detunings where combs are desired and repeat the AFC preparation procedure (see Fig. 4 in the main text for the resultant absorption profile). To achieve greater contrast, and hence more efficient AFCs \cite{afzelius2009a}, the described preparation procedure is repeated 30 times leading to an overall optical pumping duration of 5 ms. A 2 ms wait time follows the preparation -- it corresponds to 25 times the radiative lifetime of the $^3H_4$ excited level, and ensures that no luminescence masks the retrieved photons. Although a zero-phonon line of hundreds of GHz is  available in Ti:Tm:LiNbO$_3$ \cite{sinclair2010a,sun2012a}, our multimode AFC only takes advantage of approximately 10 GHz. This bandwidth is set by the sampling rate of our waveform generator and the operating bandwidth of our phase modulator. Referring to Fig. 4 in the main text, signatures of bandwidth limitations are indicated by the generation of AFCs over intervals between desired spectral bins, which is due to higher order frequency modulations, and by reduced AFC quality at larger frequency detunings, which is due to reduced energy in the first-order modulation sideband. At the end of each multimode AFC preparation sequence we sweep the laser frequency over a 200 MHz bandwidth around zero detuning. This results in widening a spectral hole (from a linewidth of a few MHz to 200 MHz) that is created during the AFC preparation procedure. This ensures that unmodulated (i.e. zeroth order) light accompanying our multimode qubits is not temporally stretched, via interaction with the spectral hole, and produce noise counts that reduce the measured storage and retrieval fidelity of qubits. \\

\section{C: Limitations to memory efficiency}

The memory efficiency (i.e. the probability for a photon to be stored and recalled) is currently approximately $1.5\times10^{-4}$.  The efficiency is determined by the transmission, or efficiency, respectively, of each element constituting our quantum memory. This includes: imperfect optical mode matching between the input and output fibres and the waveguide quantum memory, leading to a fibre-to-fibre transmission of 0.2; limited optical depth and AFC comb finesse, resulting in a probability for an incoming photon to be absorbed and re-emitted of 0.01; 50\% insertion loss into our phase modulator; average efficiency of serrodyne shifting of 0.6; imperfect mode matching into and out of the filtering cavity, resulting in additional 75\% loss.

To increase the total system efficiency, as required for a practical quantum memory, the following improvements can be made. First, better optical mode matching at both fibre-to-waveguide interfaces can raise fibre-to-fibre coupling in principle from $20$\% up to $100$\%. Alternatively, using bulk rare-earth-ion doped crystals and fibres with gradient-index lenses, transmission can also be raised to near unity. Second, the insertion loss of our  phase modulator can be overcome by integrating it with our waveguide Ti:Tm:LiNbO$_3$ crystal. Alternatively, the frequency of the recalled photons can be shifted by means of sum-frequency generation \cite{tanzilli2005a} using a pump beam of variable frequency. This can in principle be done with 100\% conversion efficiency \cite{vandevender2007a}. Third, higher-bandwidth modulators and driving electronics would allow employing less distorted serrodyne waveforms, resulting in more efficient frequency chirps. Fourth, the filtering cavity needs to be optimized for the particular application (we used an already existing cavity). Finally, in regards to the AFC itself, the quantum memory protocol theoretically allows for $100\%$ efficiency \cite{afzelius2009a}. Improvements from our current 1\% rely on increased comb contrast and comb finesse $F=\Delta / \gamma$ (where $\Delta$ is the teeth spacing and $\gamma$ is the linewidth of each tooth \cite{afzelius2009a}), and the preparation of a spatio-spectral grating \cite{tian2013a} or embedding of the rare-earth-ion-doped crystal into an impedance-matched cavity \cite{afzelius2010b,moiseev2010b,sabooni2013a,jobez2014a}. We note that the possibility to increase the finesse, as well as to achieve longer storage times, relies on minimizing the parameters $\Delta$ and $\gamma$. Since both are limited by the homogeneous linewidth of Tm, this can be achieved by, for example, reduced operating temperatures \cite{sun2012a}.\\

\section{D: Creating high finesse AFCs}

\begin{figure*}
\begin{center}
\includegraphics[width=0.9\textwidth]{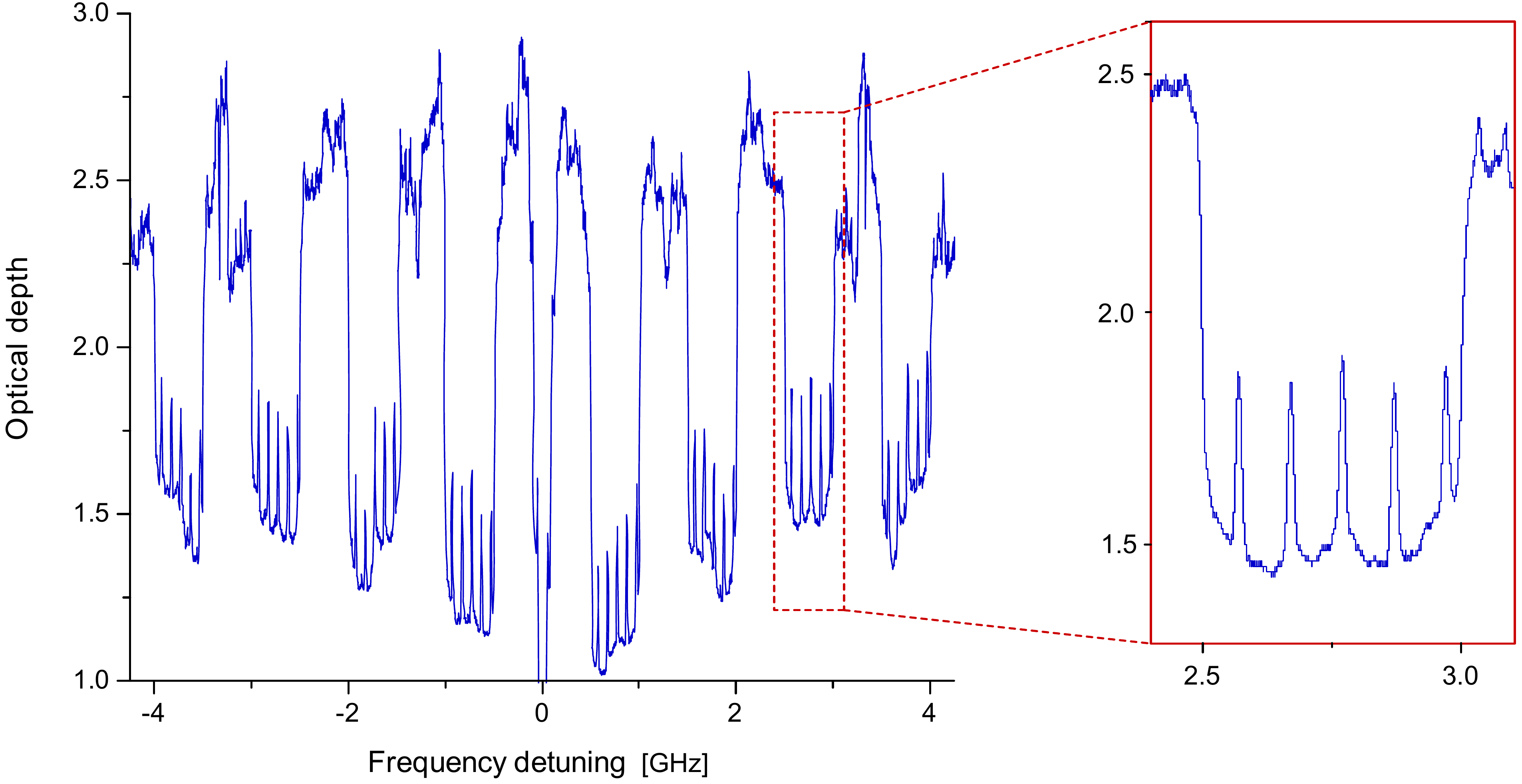}
\caption{\textbf{Multimode AFC of finesse $>$2.} Eight 500 MHz-wide AFCs, each having finesse of eight, prepared over a total bandwidth of 9 GHz.  The optical depth at the centre of each comb tooth is approximately 0.35 -- it is limited by our preparation method, Zeeman level lifetimes, and decay mechanisms present in our crystal. The reduced comb depths that are seen at larger detunings are due to reduced efficiency of the phase modulator. The magnetic field is set to 0.2 Tesla. The inset shows the AFC at the spectral bin between 2.5 and 3 GHz detuning.}
\label{fig:highfin}
\end{center}
\end{figure*}
The AFC shown in Fig. 4 of the main text is generated by optically pumping atomic population from troughs (i.e. the regions of transparency between teeth) into adjacent teeth \cite{saglamyurek2011a,jin2013a}. Towards this end, we apply a magnetic field along the C$_3$-axis of Tm:LiNbO$_3$. The ground and excited state of the $^3H_6\leftrightarrow {}^3H_4$ transition in Tm then split into two pairs of nuclear Zeeman levels \cite{sinclair2010a,sun2012a}, which are tuned such that the frequency difference between pairs is $\Delta/2$. While this approach maximizes the probability for a photon to be absorbed (all atoms continue to contribute to the absorption), it also results in the creation of an AFC with finesse of two. Assuming the use of an impedance-matched cavity \cite{afzelius2010b,moiseev2010b,sabooni2013a}, the memory efficiency $\eta$ -- in the limit of the AFC optical depth to finesse ratio being much lower than one -- is given by
\begin{equation}
\eta=e^{-\tfrac{7}{F^2}} \quad ,
\label{afc}
\end{equation}
\noindent
which limits the efficiency for an $F=2$ AFC to $17$\%. Note that this equation assumes each AFC tooth has a Gaussian shape\cite{afzelius2010b}, and ignores decay between nuclear Zeeman levels (causing loss). The latter can be approximated when operating at sufficiently low temperatures \cite{sun2012a}.

Since improvements in efficiency rely on increased comb finesse $F$, we create and measure a multimode AFC having $F=8$ in a separate experiment (see Fig.~\ref{fig:highfin}). Towards this end, we increase the magnetic field to 0.2 Tesla, and hence increase the difference in the aforementioned ground and excited-state Zeeman splitting to more than the separation between troughs and neighboring peaks. This allows optical pumping of Tm ions out of the bandwidth occupied by an AFC, and potentially a larger memory efficiency. However, as briefly described above, this also results in a reduction of effective optical depth. Yet, if an impedance-matched cavity is used, according to Eq.~\ref{afc}, this results in an efficiency of $\eta_{F=8}=90\%$.\\

\section{E: Generation of qubits in many spectral modes}
Multimode time-bin qubits are generated by intensity and frequency modulating laser light using the aforementioned AOM and phase modulator. Each qubit consists of one (early or late), or two (early and late), 15 ns-long Gaussian-shaped pulses, generated using the AOM and separated by 20 ns, with each pulse having a frequency bandwidth matching that of an AFC (i.e. a spectral bin). The fibre frequency of each qubit is defined by driving the phase modulator with a sinusoidal signal, whose frequency (more precisely: the positive-detuned first order sideband) defines that of the spectral mode to be occupied by the qubit. Furthermore, the sinusoidal signal's phase can be changed in between the generation of the first and second temporal mode, which allows creating various qubit states. Multimode qubit generation is accomplished by simultaneously driving the phase modulator with many independent sinusoidal signals. To generate the qubits shown in Fig. 4 of the main text, each sinusoidal signal was used to generating two qubits in spectral bins with with opposite-signed detunings.\\ 

\section{F: Measurement of fidelity}
The fidelity $\mathcal{F}$ quantifies how close a recalled quantum state is with respect to the originally created state. In our experiment, we employ time-bin qubits in states $\ket{\psi}\in [\ket{e}, \ket{l}, \ket{+}, \ket{-}]$, where $\ket{\pm}\equiv\frac{1}{\sqrt{2}}(\ket{e}\pm\ket{l})$. Because we employ the decoy state analysis - described in the next section - it is possible to speak of qubit states encoded into the attenuated laser pulses that we use in our experiments. The average fidelity is quantified by $\mathcal{F}= \frac{1}{3}\mathcal{F}_{e/l}+\frac{2}{3}\mathcal{F}_{+/-}$ where, for example, $\mathcal{F}_{e/l}=\frac{1}{2}(\mathcal{F}_{e}+\mathcal{F}_{l})$ and $\mathcal{F}_{e(l)}$ is the fidelity of an $|e\rangle$ ($|l\rangle$) state. The parameter $\mathcal{F}_{l}$ is calculated as $\mathcal{F}_{l}=C_{l|l}/(C_{e|l}+C_{l|l})$ where, for example, $C_{e|l}$ denotes the number of early detection events given $\ket{l}$ was originally encoded. The parameters $\mathcal{F}_{e}$, $\mathcal{F}_{+}$, and $\mathcal{F}_{-}$ are calculated in a similar way. The fidelity of time-bin qubits prepared in $\ket{e}$ or $\ket{l}$ are measured by recording photon arrival times using single photon detectors based on Si-avalanche photodiodes. Measuring qubits prepared in $\ket{+}$ or $\ket{-}$ requires detecting photons that have passed through an interferometer. This measurement as well as steps to stabilize and phase-align the interferometer are discussed next.

First, preceding each measurement, the phase of a fibre-based Mach-Zehnder interferometer having 4 m path-length difference (corresponding to 20 ns travel-time difference) and 98\% intrinsic visibility is aligned to allow projection measurements onto $\frac{1}{\sqrt{2}}(\ket{e}+\ket{l})$. Specifically, qubits prepared in $\frac{1}{\sqrt{2}}(\ket{e}+\ket{l})$, each having mean photon number of 20, are sent into the interferometer and the resultant interference signal at one particular output of the interferometer is maximized by stretching the fibre in one arm of the interferometer using a piezo actuator. To keep the phase stable, we then send light from our continuous-wave laser into the interferometer, detect the interference signal at the output of the interferometer using a linear photodetector (shown as 'D' in Fig. 3 of the main text), and keep the signal constant using feedback to the piezo actuator. This feedback is applied during each AFC preparation sequence, i.e. for 2 ms every 12 ms, which is sufficient to avoid thermal drifts. The interferometer stabilization setup is indicated by ''to (from) MZI lock setup'' in Fig. 3 of the main text.

After these steps, the measurement sequence begins by reducing the mean photon number per qubit to the single-photon level (either 0.5, 0.1, or zero photons per qubit), creating and storing photons in the desired quantum states (either $\ket{+}$ or $\ket{-}$), recalling photons from the AFC memory, and subsequently projecting them onto $\ket{+}$ and $\ket{-}$. The latter corresponds to detections in the interferometer output mentioned above, or detections in this output after having introduced an additional $\pi$ phase shift using the interferometer's piezo actuator. 

Finally, we note that frequency shifting of time-bin qubits by an amount $\Delta\nu$ comes along with a modification of the phase $\varphi$ that determines the superposition of early and late temporal modes: $\Delta\varphi=2\pi\Delta\nu\Delta t$, where $\Delta t$ denotes the difference between early and late. We use the phase modulator that preceds the filtering cavity to both frequency shift and correct any phase shifts introduced when recalling qubits.\\ 

\section{G: Bounding the single-photon fidelity using decoy state analysis}

Decoy state analysis allows implementing quantum key distribution protocols using phase-randomized attenuated laser pulses without compromising the protocol's security due to photon number splitting attacks \cite{dusek1999a,brassard2000a}. It ensures that the final key stems only from attenuated laser pulses containing one photon. The application of this analysis can be further generalized to allow any experiment performed with weak coherent states, such as attenuated laser pulses, to mimic an experiment using a much more elaborate setup that incorporates single photon (Fock) states. For the purpose of proving the quantum nature of e.g. memories, as in our experiment, the decoy state analysis provides an avenue for simpler experimental demonstrations of complex protocols. Here we provide a brief outline of how we employ the decoy state analysis to our experiment. It follows the original work by Ma \textit{et al.} \cite{ma2005a}.

In order to show the quantum nature of our spectrally multimode memory, we must verify that the fidelity of the recalled state with the input state is higher than the classical bound of 2/3 \cite{massar1995a}. However, this bound is only valid when using genuine qubits, i.e. for quantum states encoded into single photons. When using attenuated laser pulses, the classical bound has to be increased to account for the statistical distribution of the number of photons in a pulse and is also impacted by the memory efficiency \cite{specht2011a,gundogan2012a}. To be able to apply the 2/3 classical bound, we use a decoy state method to find a lower bound for the fidelity for the single photon component of the attenuated laser pulse. The derivation assumes a Poissonian photon number distribution in the light such as in the attenuated laser pulses at our memory's input. Furthermore, the pulses need to be phase randomized, which results in a states described by
\begin{equation}
\rho=\sum_{n=0}^{\infty}P(n)\proj{n}{n}\quad ,
\end{equation}
where $P(n)$ is the Poissonian distribution and $\proj{n}{n}$ denotes an n-photon Fock state.

As a first step we define the error rate 
\begin{equation}\label{eq:qberdef}
	E_{\psi}=C_{\psi_\perp|\psi}/(C_{\psi_\perp|\psi}+C_{\psi|\psi}) \quad,
\end{equation} 
where, as in the preceding section, $C_{\psi_\perp|\psi}$ denotes the number of detection events corresponding to the state $\ket{\psi_\perp}$ given the state $\ket{\psi}$ was originally encoded. Since $\ket{\psi_\perp}$ corresponds to a state orthogonal to $\ket{\psi}$, a count of this type constitutes an error. Comparing to the expression for the fidelity it is furthermore clear that $\mathcal{F}=1-E$.

Using Eq.~(25) from \cite{ma2005a} the error rate $E^{(1)}$ for the single photon component of the coherent pulses is upper bounded by $E_U^{(1)}$, which is given by
\begin{align}\label{eq:errbound}
	E^{(1)} \leq E_U^{(1)}&=\frac{E^{(\mu_{d1})}Q^{(\mu_{d1})}e^{\mu_{d1}} - E^{(\mu_{d2})}Q^{(\mu_{d2})}e^{\mu_{d2}}}{(\mu_{d1}-\mu_{d2}) Y_L^{(1)}} \nonumber \\
	&=\frac{E^{(\mu_{d1})}Q^{(\mu_{d1})}e^{\mu_{d1}} - E^{(0)}Y^{(0)}}{\mu_{d1} Y_L^{(1)}} ,
\end{align}
where $Y^{(0)}$ and $Y_L^{(1)}$ are the zero-photon yield and the lower bound for the single photon yield (defined below), respectively, $\mu_{d1}=0.1$ and $\mu_{d2}=0$ are mean photon numbers for the two \emph{decoy states} used in our experiment, and $E^{(\mu_{d1})}$ and $E^{(\mu_{d2})}=E^{(0)}$ are the corresponding error rates, which can be estimated from measurements using Eq.~\eqref{eq:qberdef}. (The second line of Eq.~\eqref{eq:errbound}} specifically assumes that $\mu_{d2}=0$, i.e. the second decoy state is a vacuum state.)
The gain $Q^{(\mu)}$ is the probability that a detector registers a count and depends on the mean photon number $\mu$ at the (memory) input and the loss up until the detector including the detector's quantum efficiency. Hence the gain can be directly calculated from the total number of counts accrued and the repetition rate of the pulses.

For phase randomized coherent states with a Poissonian photon number distribution the gain can be expressed as
\begin{equation}
	Q^{(\mu)} = \sum_{n=0}^{\infty} Y^{(n)} \frac{\mu^n}{n\mathrm{!}}e^{-\mu} \quad , 
\end{equation}
where the yield of an $n$-photon state $Y^{(n)}$ denotes the conditional probability of a detection given that an $n$-photon state was sent. The yields can generally not be directly measured without sources of photon number (Fock) states. An exception is the yield of the vacuum state, for which $Y^{(0)}=Q^{(0)}$. This fact is used to simplify Eq.~\eqref{eq:yieldbound} and in the second line of Eq.~\eqref{eq:errbound}. Instead one can derive a lower bound $Y_L^{(1)}$ for the single photon yield (used in Eq.~\eqref{eq:errbound}), which for the specific case of $\mu_{d2}=0$ is given by \cite{lucio2009a}
\begin{align}\label{eq:yieldbound}
	Y^{(1)} \geq Y_L^{(1)} =&
	\frac{\mu_{s}}{\mu_{s}\mu_{d1}-\mu_{d1}^2} \bigg( Q^{(\mu_{d1})}e^{\mu_{d1}} \qquad \\
	&\qquad \quad -\frac{\mu_{d1}^2}{\mu_{s}^2}Q^{(\mu_{s})}e^{\mu_{s}} - \frac{\mu_{s}^2-\mu_{d1}^2}{\mu_{s}^2}Y^{(0)} \bigg)  \nonumber
\end{align}
where $\mu_{s}=0.5$ is the mean photon number of the \emph{signal state}. The right-hand sides of equations \eqref{eq:errbound} and \eqref{eq:yieldbound} now contain directly measurable values and thus allow us to calculate the upper bound on the error rate $E_U^{(1)}$. By means of Eq.~\eqref{eq:qberdef} we can compute the lower bound on the fidelity
\begin{equation}
	\mathcal{F}^{(1)}=1-E^{(1)}\geq1-E_U^{(1)}\equiv \mathcal{F}_L^{(1)} \ ,
\end{equation}
which thus allows us to calculate the values in Table~I in the main text.

This covers the essence of the decoy state analysis. Using it we derive a bound on the fidelity that we would have achieved if -- all other things the same -- we had utilized true single photons to encode qubits at the memory input.
\begin{figure}[h!]
		\includegraphics[width=\columnwidth]{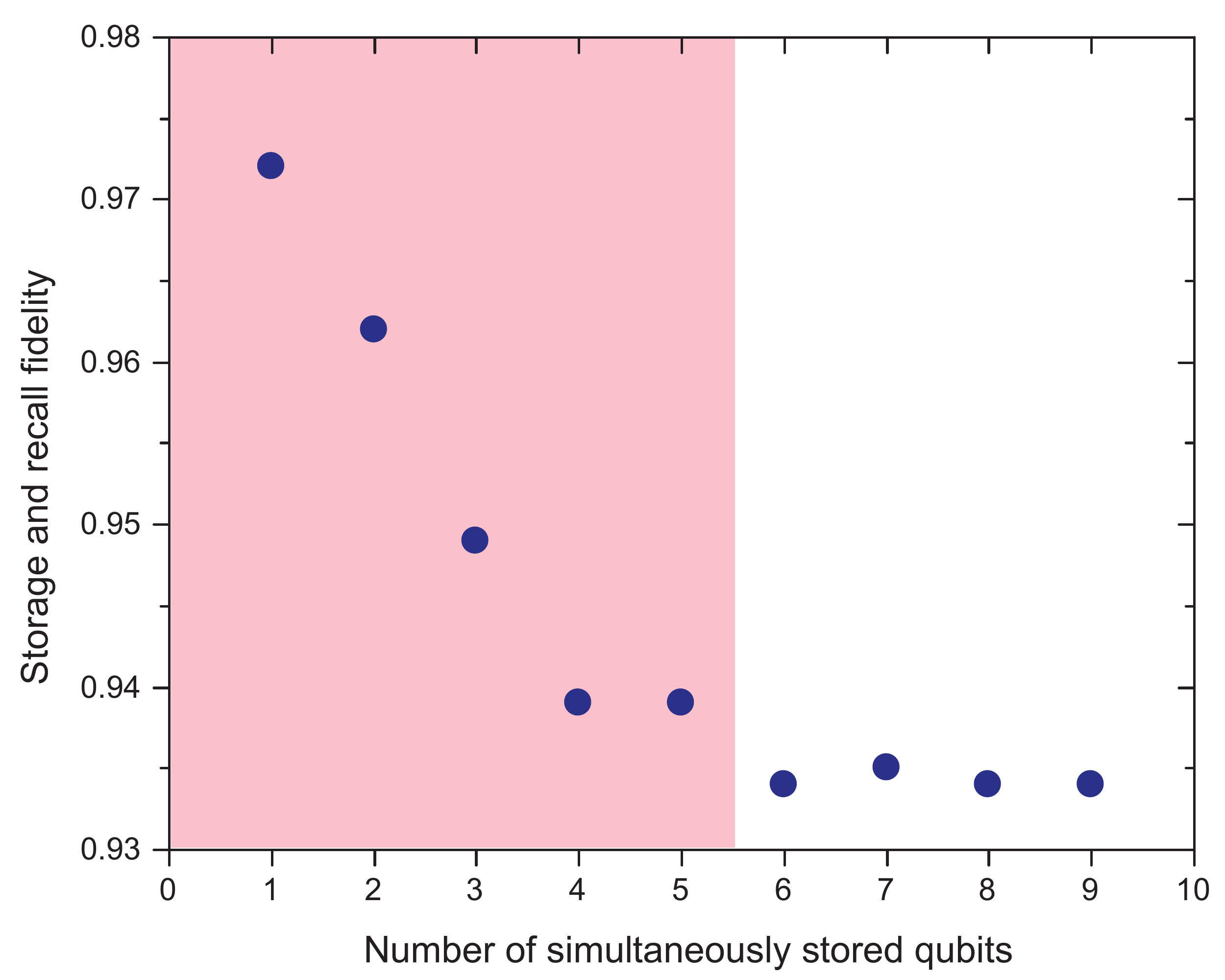}
		\caption{\textbf{Measurement of cross-talk.} Dependence of storage and recall fidelity of the test qubit on the number of simultaneously stored qubits in neighboring spectral bins. The test qubit in $|l\rangle$ occupied the spectral bin having 1350 MHz detuning, while orthogonal qubits were added one-by-one to neighboring spectral bins in the following order: 1650 MHz, 1050 MHz, 1950 MHz, 750 MHz, etc. The cavity resonance was set to 2.85 GHz detuning. We find that the fidelity is constant when storing more than 5 modes simultaneously (shown in the white region of the plot). Hence, crosstalk (due to the Lorenzian linewidth of the cavity) is limited to coming from the nearest and second-nearest neighbour (shown in the shaded region of the plot). A further reduction in fidelity is due to limited frequency shift efficiency of our phase modulator, and is independent of the number of qubits simultaneously stored. Each projection measurement was taken over 60 s, the mean photon number per qubit was 0.6, and uncertainty (one standard deviation) was calculated from error propagation using statistical uncertainties of photon counts (not shown as it is smaller than the symbol size).}
\label{fig:fidvsmodes}
\end{figure}

\section{H: Measurement of cross-talk between spectral modes}
To examine the effect of cross-talk between spectral modes, we first store and retrieve a 'test' qubit prepared in $\ket{l}$ in the spectral bin having 1350 MHz detuning (refer to Fig. 4 of the main text). We then shift the test qubit into cavity resonance having 2.85 GHz detuning, measure the probability to project it onto $\ket{l}$ and calculate $\mathcal{F}_{l}$. We then increase the number of simultaneously stored qubits by creating them in neighbouring spectral bins, and repeat the fidelity measurement with the test qubit. Note that all additional qubits are prepared in the orthogonal $\ket{e}$ state, such that the reduction of the fidelity of the test qubit due to cross-talk is maximized. The results are shown in Figure \ref{fig:fidvsmodes}. We find that while there is a small amount of cross-talk between neighbouring modes, it is only significant when considering qubits separated by at most two spectral bins. Improvements in the fidelity can be achieved, for example, by increasing the separation between spectral bins, or by employing a filtering cavity with a steeper transition from resonance.\\

\end{document}